\documentclass[lettersize.journal]{IEEEtran}
\usepackage{graphicx} 
\usepackage{amsmath,amssymb,amsfonts}
\usepackage{algorithmic}
\usepackage{textcomp}
\usepackage{xcolor}
\usepackage[table]{xcolor}
\usepackage{tabulary}
\newcolumntype{K}[1]{>{\centering\arraybackslash}p{#1}}
\usepackage{array}
\usepackage{url}
\usepackage{authblk}
\usepackage{multirow}
\usepackage[sorting = none, backend = bibtex, style=numeric-comp]{biblatex}
\usepackage{caption}
\begin{document}

\title{Performance analysis of classical adiabatic annealing on Ising machines}

\author[1*]{Jacob Lamers}
\author[]{Guy Verschaffelt}
\author[]{Guy Van der Sande}

\affil[]{Applied Physics Research Group, Vrije Universiteit Brussel, Pleinlaan 2, 1050 Brussels, Belgium}

\maketitle

\begin{abstract}
Ising machines are a promising approach to solve combinatorial optimization problems. They map these problems onto the Ising model and search for low-energy configurations. However, navigating the rugged energy landscapes of these systems remains difficult. To improve this navigation, classical adiabatic annealing has been proposed in the literature as a heuristic optimization method for classical Ising machines. Using this technique, the Hamiltonian of the Ising machine is gradually transformed from an easily solvable Hamiltonian to the target Hamiltonian. However, its purported effectiveness is primarily motivated by an analogy to quantum adiabatic annealing, and systematic benchmarking has remained limited. 

In this work, we analyze the classical adiabatic annealing technique using continuation methods. Motivated by insights from this analysis, we propose an optimized annealing strategy we refer to as hybrid classical adiabatic annealing. We benchmark our proposed strategy using MaxCut instances with up to 800 spins and problems with external fields, for which it achieves a marginal improvement for a limited set of problems. We conclude that, although theoretically motivated and occasionally beneficial, the hybrid strategy does not offer a sufficient practical advantage over simpler, existing techniques.
\end{abstract}

\section{Introduction}
Complex combinatorial optimization problems (COP) are ubiquitous in real-life applications \cite{Ising_formulations_of_many_NP_problems} ranging from finance \cite{Fincance}, power system operations \cite{ApplicationsInPowerSystemOperations}, traffic flow optimization \cite{Traffic_Flow}, route scheduling \cite{Job_Scheduling}, logistics \cite{SupplyChainLogisticsWithAnnealing} to computational biology \cite{ComputationalBiology, Protein_Folding}. Ising machines (IM) have emerged as dedicated heuristic solvers for such problems. The key idea is to map the COP onto an Ising problem, which consists of $N$ binary variables called spins $\sigma_i\in\{-1,1\}$, interacting with their neighbors. The energy of a configuration of spins is given by the Ising Hamiltonian
\begin{equation}
    \label{eq:ising_hamiltonian}
    \mathcal{H} = -\frac{1}{2}\sum_{ij}^N J_{ij}\sigma_i\sigma_j - \sum_i^N h_i\sigma_i,
\end{equation}
where $J_{ij}$ is the coupling matrix and $h_i$ are external fields. Low-energy states of this Hamiltonian correspond to near-optimal solutions of the COP, while the ground state represents the optimal solution. The goal of the IM is therefore to find low-energy states of the Ising Hamiltonian.

Many IM implementations have been proposed so far \cite{OverviewMcMahon}. Some of them use binary variables to represent the spins such as quantum annealers \cite{Quantum_annealing_DWave}, digital annealers \cite{DigitalAnnealer}, SLM based IMs \cite{Conti} and systems based on probabilistic bits (p-bits) \cite{Probabilistic_computing_with_pbits, Stochastic_pbits_for_invertible_logic, Integrated_pbits}. In this work, however, we focus on implementations that represent spins using continuous analog variables $s_i\in\mathbb{R}$. Examples include systems based on degenerate optical parametric oscillators (DOPOs) \cite{2000nodes, 100000SpinsCIM}, opto-electronic oscillators \cite{Poor_mans_CIM, 200GOPS_PhotonicIsingMachine}, electrical resonators \cite{Ising_machine_based_on_networks_of_subharmonic_electrical_resonators}, memristor crossbar arrays \cite{jiang2023efficient}, and polariton condensates \cite{PolaritonCondensates, PolaritonCondensates_2}.

The energy landscape of the problem is typically rugged with many local minima. Therefore, many strategies have been developed to avoid or escape these local minima. One such strategy is quantum adiabatic annealing, which is used by quantum annealers. This technique is based on the quantum adiabatic theorem which states that a quantum system initially prepared in the ground state of a slowly varying, gapless Hamiltonian will remain in the instantaneous ground state throughout the evolution as long as this evolution is slow enough \cite{QuantumAdiabaticAnnealingOnRandomProblems, QuantumAdiabaticAnnealingOverview}. So when the system is initialized in the known ground state of a Hamiltonian and that Hamiltonian is gradually transformed into that of the optimization problem at hand, the system is expected to reach the ground state of the problem Hamiltonian at the end of the annealing.

While quantum adiabatic annealing offers a promising route to solving combinatorial optimization problems, building a large-scale, energy-efficient quantum system with sufficient coherence to support this process remains a significant challenge. Therefore, classical analogues of this technique have been proposed, even though there is no direct classical counterpart to the quantum adiabatic theorem. In one such approach, the IM is initialized in the known ground state of a different Hamiltonian to the one that needs to be solved. This Hamiltonian is then slowly transformed to the target Hamiltonian during the run. We will refer to this approach as Classical Adiabatic Annealing (CAA) and although it has been studied before \cite{Pierangeli_AdiabaticAnnealing, CAA_Thomas}, its performance has not yet been thoroughly benchmarked and compared to other methods. 

In this work, we provide insight in the performance of CAA on classical analogue IMs, and based on this, we introduce an optimized strategy for performing CAA. To gauge its competitiveness, we compare it to a method commonly used to enhance the performance of IMs by gradually increasing either the linear gain $\alpha$ or coupling strength $\beta$ parameter during a run \cite{paper_Ganguli, PaperJacob_UsingContinuationMethods, Paper_Leen_predictingoptimalnoisestrength, inspiration_idea_thomas, Mean-field_CIM_with_artificial_ZeemanTerms}. This technique, which we will refer to as regular annealing (RA), has been shown to deliver a comparable performance to other state of the art methods such as chaotic amplitude control \cite{Paper_Leen_predictingoptimalnoisestrength}. For this comparison, we consider MaxCut instances with up to $800$ spins as well as Beasley instances that include external fields. For MaxCut problems, we find that while the hybrid CAA strategy consistently outperforms regular annealing (RA), the observed gains are modest and unlikely to justify the added complexity of the method relative to the much simpler RA. In contrast, for instances with external fields, hybrid CAA achieves a more substantial reduction in time‑to‑solution. However, prior work has demonstrated that in this regime the performance of RA can be dramatically improved by computing the spin couplings using the sign of the variables, a technique we refer to as the spin sign method \cite{ExternalFields}. When compared against this enhanced baseline, hybrid CAA loses its advantage.

\section{Results}
\label{sec:Results}
In analog Ising machines, the spins of the Ising model are replaced by analog variables referred to as spin amplitudes $s_i$. The time evolution of these spin amplitudes can be described by a set of differential equations, such as e.g.
\begin{multline}
    \label{eq:tanh_update}
    \frac{ds_i}{dt} = -s_i\\  + \tanh\left[\alpha s_i + \beta\left(\sum_j^N J_{ij}s_j + h_i\right)+\gamma\xi_i(t)\right],
\end{multline}
where $\alpha$ is the linear gain, $\beta$ is the interaction strength and $\gamma$ is the noise strength parameter. $\xi$ represents Gaussian white noise at time $t$. The hyperbolic tangent nonlinearity is chosen here for its superior performance over other nonlinearities typically used to describe IMs \cite{Order_of_magnitude, PaperJacob_UsingContinuationMethods}. Furthermore, this nonlinearity provides a good description of any IM featuring a pitchfork bifurcation combined with saturation of the spin amplitudes \cite{PaperJacob_UsingContinuationMethods}. To evaluate the system's energy at any point in time, the signs of the spin amplitudes are used as binary spins $\sigma_i = \text{sgn}(s_i)$, and the Hamiltonian in Eq.~\eqref{eq:ising_hamiltonian} is computed. In a typical run of the IM, the system converges to a fixed point.

To evaluate the performance of the IM, we first focus on MaxCut problems, where the task is to divide a graph in two subgraphs such that the number of connections between these two subgraphs is as large as possible. The MaxCut problem is mapped to the Ising problem by replacing each node with a spin and each edge with an antiferromagnetic coupling. This formulation therefore does not use external fields ($h_i=0$). The instances we consider are all unweighted instances of the BiqMac library with a $50\%$ connectivity ratio \cite{BiqMac} and all instances with $800$ spins of the GSET library \cite{GSET}. As the presence of the external fields $h_i$ can significantly influence the performance of the IM \cite{ExternalFields}, we also consider the Beasley instances with problem sizes up to $250$ spins of the BiqMac library \cite{BiqMac}. These are quadratic 0-1 programming problems that require external fields when formulated as an Ising problem.

The basic principle of CAA is that the coupling matrix is  gradually transformed from some initial coupling matrix $J_{ij}^{ \text{Initial}}$ to the target coupling matrix $J_{ij}^{ \text{Target}}$. Therefore, we make the coupling matrix of the Ising machine time-dependent via
\begin{equation}
    \label{eq:CouplingMatrixCAA}
    J_{ij}(t) = (1-\mathcal{F}(t))J_{ij}^{ \text{Initial}} + \mathcal{F}(t)J_{ij}^{ \text{Target}},
\end{equation}
where $\mathcal{F}(t)$ is a function of time that varies from zero to one sufficiently slowly. The initial coupling matrix $J_{ij}^{ \text{Initial}}$ represents a Hamiltonian of which the ground state is easily found. 

We have investigated several options for the initial coupling matrix $J_{ij}^\text{Initial}$, and, as explained in the appendix, initializing the system with all-to-all ferromagnetic coupling yields the best performance for the problems considered. In this case, the initial coupling matrix is an all-ones matrix with zeros on the diagonal and with the same dimensions as the target coupling matrix $J_{ij}^{ \text{Target}}$:
\begin{equation}
    \label{eq:InitialCouplingMatrix_AllToAll_Ferromagnetic}
    J_{ij}^{ \text{Initial}} = \begin{pmatrix}
        0&1&1&\dots &1\\
        1&0&1&&1\\
        1&1&0&&\vdots\\
        \vdots&&&\ddots&1\\
        1&\dots & &1 &0
    \end{pmatrix}
\end{equation}
The ground state of this system has all spins aligned. The spin amplitudes are all equal and are such that the right-hand side of Eq.~(\ref{eq:tanh_update}) is zero. They can be determined via Newtons method and depend on both system parameters $\alpha$ and $\beta$. 

\subsection{Continuation of classical adiabatic annealing}
\label{subsec:Continuation_of_classical_adiabatic_annealing}

\begin{figure*}[t]
    \centering
    \includegraphics[width=0.85\linewidth]{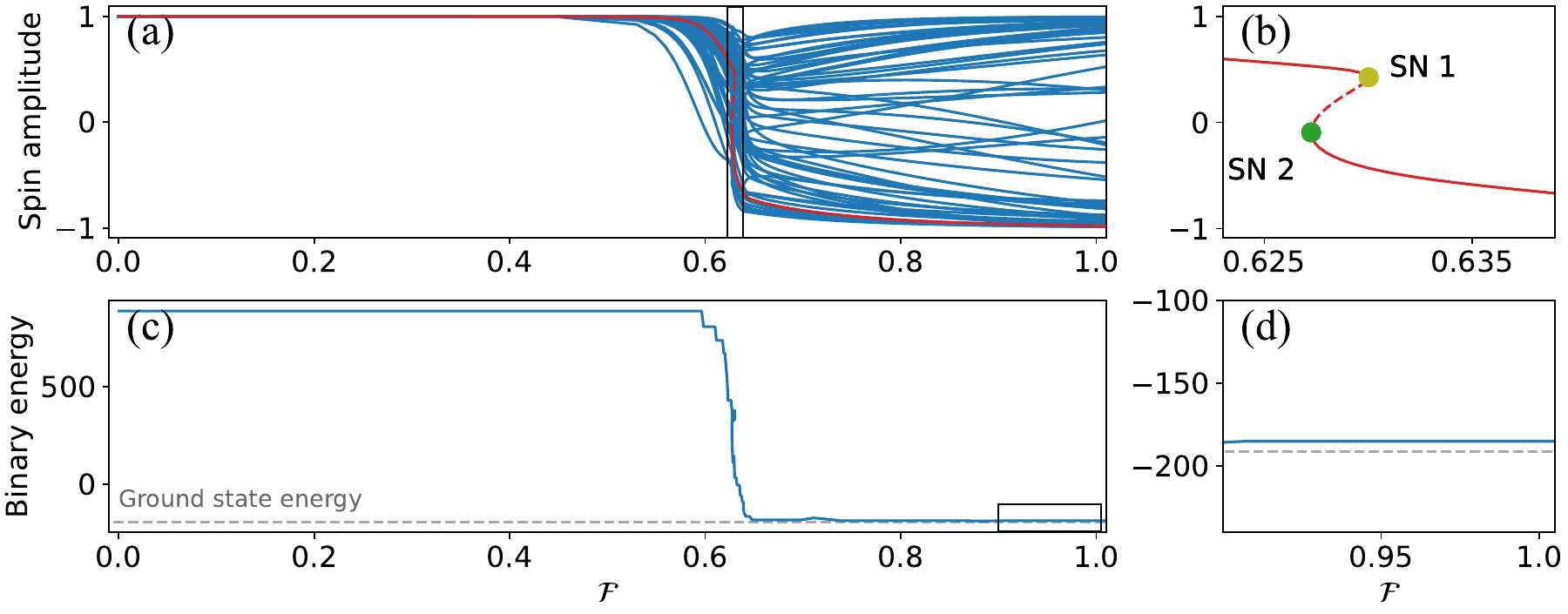}
    \caption{{\bf Typical example of the evolution under classical adiabatic annealing.} Continuation  as the interpolation parameter $\mathcal{F}$ varies from zero to one of (a) the spin amplitudes of problem g05\_60.3 for $\alpha=0$ and $\beta=0.25$, (b) spin amplitude $6$ in the black rectangle of (a), (c) the energy calculated using the target coupling matrix. (d) Zoom of the energy evolution in the black rectangle of (c). The energy of the ground state of the binary Hamiltonian is represented by the gray dashed line. Solid (dashed) lines are used when the spin amplitude is part of a
stable (unstable) state. SN (green and yellow dots) saddle node bifurcation} 
    \label{fig:TypicalAdiabaticAnnealing}
\end{figure*}

When performing CAA, $\mathcal{F}$ is gradually increased from zero to one. During this process, the fixed points of Eq.~\ref{eq:tanh_update} will change as well. If $\mathcal{F}$ varies sufficiently slowly, the IM will relax to the stable fixed point, corresponding to the new value of $\mathcal{F}$. In contrast, a rapid variation in $\mathcal{F}$ can cause the IM to deviate from this path, potentially leading it to other fixed points. For CAA to be a viable strategy, the fixed points encountered during a slow evolution of $\mathcal{F}$ should lead to a fixed point corresponding to the ground state of the target Hamiltonian. Therefore, before analyzing the impact of the rate of change of $\mathcal{F}$, it is essential to first examine whether these fixed points indeed guide the IM towards the desired ground state. For this, we perform a continuation analysis.

Continuation is a numerical technique used to track fixed points of a dynamical system as a parameter is varied. The function $\mathcal{F}$ will now be treated as a parameter, we refer to as the interpolation parameter and is now no longer a function of time. In this article, we employ the continuation package auto-07p \cite{AUTO} to study the evolution of the ground state fixed point of the initial coupling matrix as the interpolation parameter $\mathcal{F}$ transitions from $0$ to $1$. The set of fixed points obtained by tracking an initial fixed point during the evolution of $\mathcal{F}$ is referred to as a branch. If the fixed points of such a branch are all (un)stable, the branch is said to be (un)stable. The fixed points obtained in this manner correspond to the trajectory of a noiseless IM in the limit where $\mathcal{F}$ varies infinitely slowly. For finite adiabatic annealing rates, however, the IM may not have sufficient time to relax to the instantaneous fixed point corresponding to each value of $\mathcal{F}$.

To illustrate such an evolution, we use MaxCut instance g05\_60.3 from the BiqMac library \cite{BiqMac} as a typical example. Fig.~\ref{fig:TypicalAdiabaticAnnealing}(a) illustrates how all spin amplitudes of the fixed point representing the ground state of the initial Hamiltonian evolve as $\mathcal{F}$ increases from 0 to 1. At $\mathcal{F} = 0$, the coupling matrix is the initial coupling matrix and the system starts in a state where all spins are aligned and have the same amplitudes. The continuation algorithm then follows this fixed point as $\mathcal{F}$ increases, ultimately reaching a fixed point of the target Hamiltonian at $\mathcal{F} = 1$. The corresponding energy evolution, calculated using the target coupling matrix is shown in Fig.~\ref{fig:TypicalAdiabaticAnnealing}(c). A close up of this energy, shown in Fig.~\ref{fig:TypicalAdiabaticAnnealing}(d), reveals that the energy of the final state obtained via continuation, in this example, does not coincide with the ground state energy of the target Hamiltonian, marked by the gray dashed line. Therefore, this method is not guaranteed to lead to the GS solution.

Fig.~\ref{fig:TypicalAdiabaticAnnealing}(a) shows the complete continuation, but in order to more clearly show what is going on, we focus on spin amplitude number 6 in Fig.~\ref{fig:TypicalAdiabaticAnnealing}(b). This particular spin amplitude is chosen because it most clearly illustrates the relevant changes. Fig.~\ref{fig:TypicalAdiabaticAnnealing}(b) shows that the stable branch terminates in a saddle-node bifurcation (SN 1). The associated unstable branch loops back toward lower values of $\mathcal{F}$ and eventually merges with a different stable branch at a second saddle-node bifurcation (SN 2). This new stable branch persists up to $\mathcal{F} = 1$. We do note that bifurcations are changes to fixed points, not just spin amplitude 6.

The presence of these saddle-node bifurcations (SNs) can have detrimental consequences. During CAA, the IM is initialized in the state where all spins are in the up-state and the evolution of the IM follows Eq.~(\ref{eq:tanh_update}). For the initial value of $\mathcal{F}$, this state is a fixed point. As $\mathcal{F}$ is increased, the fixed point is altered, and the IM will momentarily not be exactly at this fixed point. However, if CAA is performed sufficiently slowly, the changes of the fixed point are small, and in the absence of noise the IM will reliably relax to the next fixed point on the branch. Consequently, when performing slow, noise-free CAA, the IM will track the stable branch shown in Fig.~\ref{fig:TypicalAdiabaticAnnealing}(a) as $\mathcal{F}$ increases. However, when the first SN is encountered, the fixed point disappears, causing the IM to transition to another nearby fixed point. Importantly, there is no guarantee that this new fixed point is part of the other stable branch shown in Fig.~\ref{fig:TypicalAdiabaticAnnealing}(b). Thus, the presence of such saddle-node bifurcations can be detrimental, as they can lead the system away from the desired solution branch. In the next section, we will show that for this example benchmark problem, both the failure to reach the ground state and the presence of SNs can be remedied. 

\subsection{Improving classical adiabatic annealing}

\begin{figure}[t]
    \centering
    \includegraphics[width=0.99\linewidth]{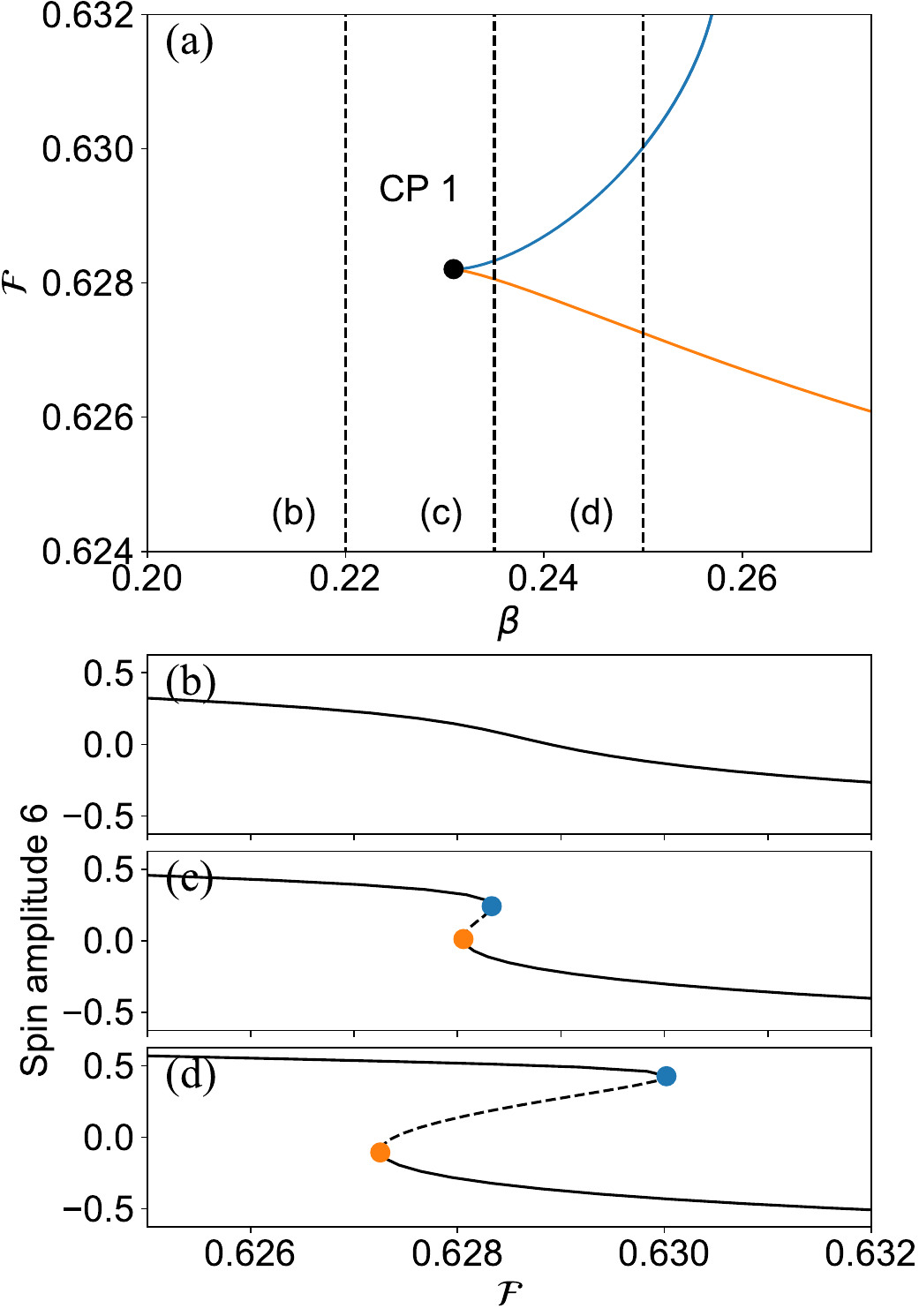}
    \caption{{\bf Positions of the saddle node bifurcations for problem g05\_60.3}(a) The value of the interpolation parameter $\mathcal{F}$ of the two saddle node bifurcations (SNs) as a function of the coupling strength $\beta$. The Cusp point is indicated by a black dot, annotated with CP 1. (b)-(d) The evolution of spin amplitude 6, used as a proxy of the fixed point, as a function of $\mathcal{F}$ for three different fixed $\beta$ values ($0.22$, $0.235$ and $0.25$). These $\beta$ values are indicated by the black dashed lines in (a). The blue and orange dots represent the SNs. These results were obtained using a constant linear gain $\alpha=0$.}
    \label{fig:3RegionsCusp_g05_60.3}
\end{figure}

We start by investigating how the values of $\mathcal{F}$ at which the SNs occur ($\mathcal{F}_\text{SN}$) change when varying the coupling strength $\beta$. Fig.~\ref{fig:3RegionsCusp_g05_60.3}(a) shows these $\mathcal{F}_\text{SN}$ values as a function of the coupling strength $\beta$. To illustrate this more clearly, Figs.~\ref{fig:3RegionsCusp_g05_60.3}(b)–(d) depict the evolution of spin amplitude 6, again used as a proxy for the fixed point as a whole, for three fixed values of $\beta$, indicated by the black dashed lines in panel (a). Fig.~\ref{fig:3RegionsCusp_g05_60.3}(d) is obtained at a fixed $\beta=0.25$, so this panel depicts the same evolution as Fig~\ref{fig:TypicalAdiabaticAnnealing}(b). As $\beta$ decreases, the $\mathcal{F}$ values of the SNs move closer together, as illustrated in Fig~\ref{fig:3RegionsCusp_g05_60.3}(c). The SNs eventually merge at a cusp point (CP) when $\beta = 0.23$, indicated by CP 1 in panel (a). For $\beta$ values below this threshold, the branch no longer contains any SNs as shown in Fig~\ref{fig:3RegionsCusp_g05_60.3}(b). This implies that if CAA is performed slowly enough and without noise for this value of $\beta$, the IM will be able to follow this branch until $\mathcal{F}=1$. 

We have repeated this analysis for all other MaxCut instances of the BiqMac library and observed that decreasing $\beta$ removes the SNs. So, during CAA, $\beta$ should be as small as possible in order to avoid SNs. However, $\beta$ should not be set too low and should remain above a critical value related to a pitchfork bifurcation of the origin (the state with all spin amplitudes equal to zero). If $\beta$ drops below this critical value, the origin becomes the only stable state, and all spin amplitudes will collapse to zero, effectively losing all progress. This critical $\beta$ value is given by  \cite{PaperJacob_UsingContinuationMethods}
\begin{equation}
    \beta_\text{crit} = \frac{1-\alpha}{\mu_1},
\end{equation}
where $\mu_1$ is the largest eigenvalue of the coupling matrix $J$. As $J_{ij}$ depends on $\mathcal{F}$ when employing CAA, $\beta_\text{crit}$ is also a function of $\mathcal{F}$. Fig.~\ref{fig:PB_Beta_AdiabaticAnnealing} shows the evolution of $\beta_\text{crit}$ as a function of $\mathcal{F}$ for problem instance g05\_60.3. At $\mathcal{F}=0$, $\beta_\text{crit}=0.0169$. As $\mathcal{F}$ increases, $\beta_\text{crit}$ becomes larger, reaching a maximum around $\mathcal{F} = 0.65$. To ensure the system remains above the critical threshold throughout CAA, $\beta$ must always remain larger than the maximum value of $\beta_\text{crit}$ encountered during the annealing. Accordingly, we enforce an arbitrary distance of $0.02$ above the maximum value of $\beta_\text{crit}$:
\begin{equation}
\label{eq:beta_min}
\beta_{\text{min}} = \max_{\mathcal{F}} \left( \beta_\text{crit} \right) + 0.02.
\end{equation}
This choice ensures that the CAA is performed with a value of $\beta$ that is as low as possible while avoiding dropping below the critical $\beta$ value.

The presence of such a critical $\beta$ value for MaxCut instances is a consequence of the symmetry between states that differ by a global minus sign. As the external fields break this symmetry, $\beta$ can be chosen arbitrarily small for problems with external fields.


\begin{figure}
    \centering
    \includegraphics[width=0.99\linewidth]{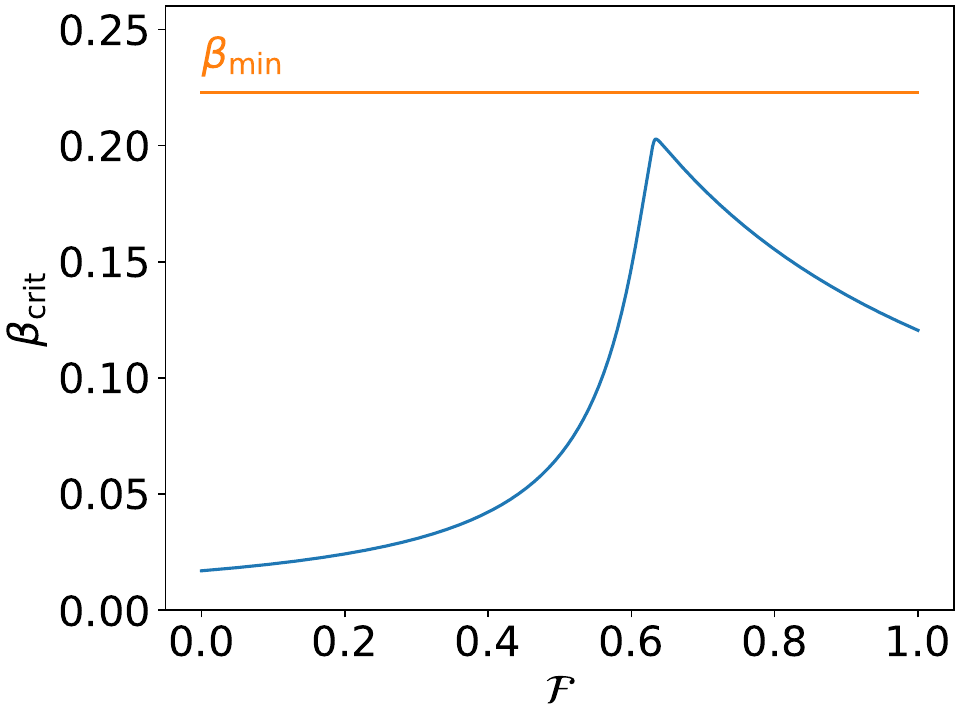}
    \caption{{\bf Critical $\boldsymbol{\beta}$ values for problem g05\_60.3.} $\beta_\text{crit}$ as a function of the interpolation parameter $\mathcal{F}$ for $\alpha = 0$. $\beta_{\text{min}}$ is defined by adding $0.02$ to the maximum $\beta_\text{crit}$ value over all $\mathcal{F}$ values and is indicated by the orange line.}
    \label{fig:PB_Beta_AdiabaticAnnealing}
\end{figure}


\begin{figure*}
    \centering
    \includegraphics[width=0.85\linewidth]{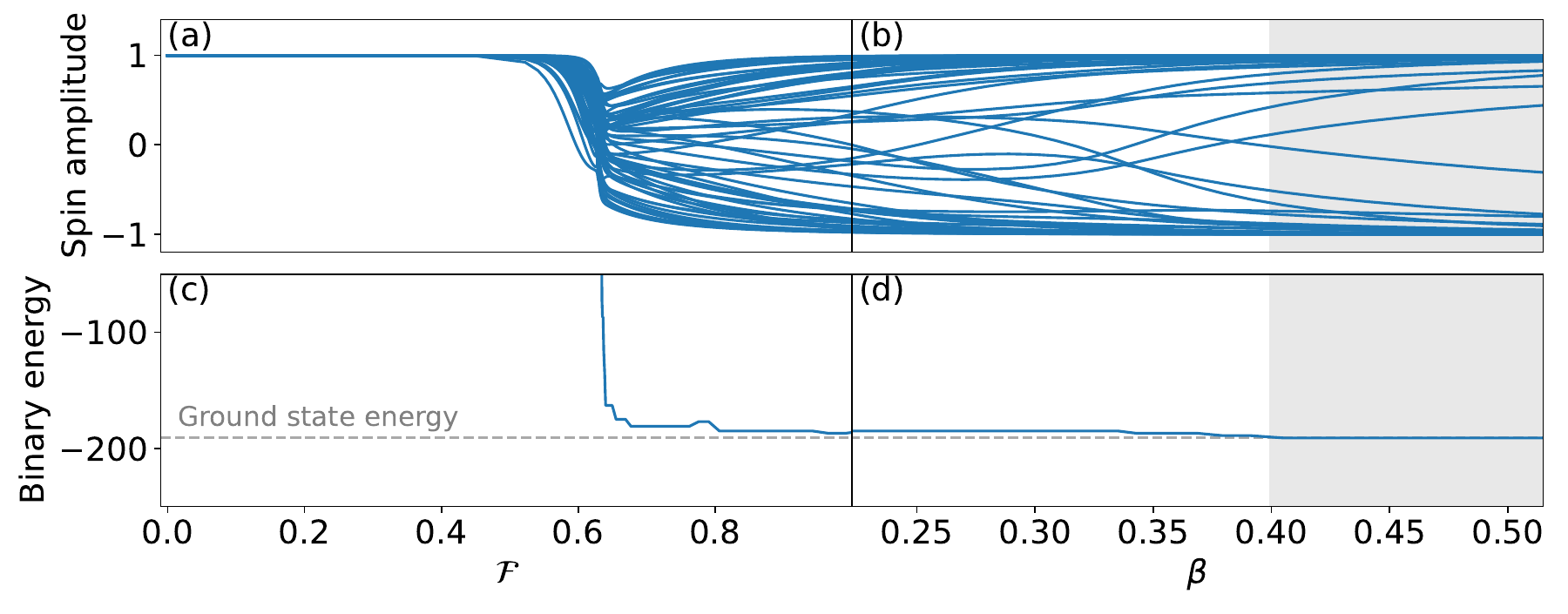}
    \caption{{\bf Continuation of CAA using $\boldsymbol{\beta_{\text{min}}}$ followed by RA for problem g05\_60.3 with $\boldsymbol{\alpha = 0}$.}(a): evolution of all spin amplitudes as the interpolation parameter $\mathcal{F}$ is increased from zero to one for a fixed $\beta=\beta_{\text{min}}=0.222$. (b) application of RA to the final state of panel (a). The energy calculated using the target coupling matrix for both (c) CAA and (d) RA. The energy of the ground state of the binary Hamiltonian is represented by the gray dashed line. The region where the fixed point corresponds to this ground state is indicated by the gray background.}
    \label{fig:CAA_RA_g05_60.3}
\end{figure*}

\subsection{Hybrid classical adiabatic annealing}

Performing CAA at such low values of $\beta$ introduces an additional complication. A fixed point of the IM is guaranteed to correspond to a minimum of the Ising Hamiltonian in Eq.~\ref{eq:ising_hamiltonian} only in the limit $\beta \rightarrow \infty$ \cite{paper_Ganguli}. So, at small $\beta$ values, the fixed point reached at the end of the CAA does not necessarily map to a minimum of the Ising Hamiltonian. Prior work has shown that this issue can be resolved by gradually increasing $\beta$ using RA. This procedure tracks the fixed point to higher $\beta$ values. 

Based on these insights, we propose a two-stage process we refer to as hybrid CAA. First, we perform CAA using $\beta = \beta_{\text{min}}$, yielding a fixed point of the target problem for $\beta_{\text{min}}$. Next, we perform RA on this state while keeping $\mathcal{F}=1$ fixed, to gradually increase $\beta$ until it is large enough for the fixed point to correctly map to a minimum of the Ising Hamiltonian. Fig.~\ref{fig:CAA_RA_g05_60.3} illustrates the result of this method for problem instance g05\_60.3. Fig.~\ref{fig:CAA_RA_g05_60.3}(a) shows the evolution of all spin amplitudes as a function of $\mathcal{F}$, similar to Fig.~\ref{fig:TypicalAdiabaticAnnealing}(a), but using $\beta=\beta_{\text{min}}=0.222$ instead of $\beta=0.25$. As a result, the branch in Fig.~\ref{fig:CAA_RA_g05_60.3}(a) does not have any SNs. At $\mathcal{F}=1$, the state is a fixed point of the target problem instance at $\beta=\beta_{\text{min}}=0.222$. Panel (b) shows the subsequent RA phase, where the coupling strength $\beta$ is slowly increased, starting from $\beta_{\text{min}}$, while keeping $\mathcal{F}$ fixed at $1$. Figs.~\ref{fig:CAA_RA_g05_60.3}(c) and (d) show the corresponding binary energy obtained using the target coupling matrix $J_{ij}^{ \text{Target}}$.  The gray dashed line represents the ground state energy of the binary Hamiltonian. As shown in Fig.~\ref{fig:CAA_RA_g05_60.3}(c), the fixed point obtained at the end of CAA does not yet correspond to the ground state. Only after $\beta$ is increased above $0.4$, highlighted by the gray background in panel (d), does the fixed point represent the correct ground state. This hybrid approach, combining CAA and RA, therefore enables us to avoid SNs by using a sufficiently low initial $\beta$, while ultimately converging to a fixed point at a large enough $\beta$ such that it faithfully represents a minimum of the binary problem \cite{paper_Ganguli, PaperJacob_UsingContinuationMethods}.

Continuation of this hybrid CAA method, CAA using $\beta_{\text{min}}$ followed by RA, is applied to all unweighted instances of the BiqMac library with 50\% edge density. For each instance, this procedure is repeated for the following values of the linear gain parameter $\alpha$: $0.7$, $0.5$, $0$, $-1$, $-2$, $-5$. If, for any value of $\alpha$, the continuation successfully reaches the ground state of the target Hamiltonian without encountering any SNs, the problem is classified as adiabatic easy, and the corresponding entry in the first column of Table \ref{tab:AdiabaticEasy} is marked in green. Otherwise, the problem is called here adiabatic hard and the entry is marked in red. If a problem is adiabatic easy, performing noise-free and infinitely slow hybrid CAA is guaranteed to result in the ground state of the Ising Hamiltonian. Conversely, if a problem is adiabatic hard, performing this idealized hybrid CAA is not guaranteed to reach the ground state. 

Similarly, if the RA continuation reaches the ground state for any of the $\alpha$ values without encountering SNs, the instance is labeled Ising easy, following the terminology of Ref.~\cite{PaperJacob_UsingContinuationMethods}, and the corresponding entry in the second column of Table~\ref{tab:AdiabaticEasy} is marked in green. Otherwise, the instance is considered Ising hard and marked in red.

\begin{table}[]
    \caption{{\bf Classification of all problem instances from the BiqMac library}. Ising or adiabatic easy  problem instances are indicated in green and Ising or adiabatic hard in red.}
    \centering
    \begin{tabular}{|K{2.5cm}|K{2.5cm}|K{2.5cm}|}
        \hline
    Problem name & Hybrid CAA & RA \\
        \hline
        \hline
        g05\_60.0 & \cellcolor[HTML]{65E14C} & \cellcolor[HTML]{65E14C} \\
        g05\_60.1 & \cellcolor[HTML]{65E14C} & \cellcolor[HTML]{65E14C} \\
        g05\_60.2 & \cellcolor[HTML]{65E14C}& \cellcolor[HTML]{65E14C} \\
        g05\_60.3 & \cellcolor[HTML]{65E14C} & \cellcolor[HTML]{65E14C} \\
        g05\_60.4 & \cellcolor[HTML]{65E14C} &\cellcolor[HTML]{65E14C} \\
        g05\_60.5 & \cellcolor[HTML]{65E14C} & \cellcolor[HTML]{65E14C} \\
        g05\_60.6 &\cellcolor[HTML]{65E14C} &\cellcolor[HTML]{65E14C} \\
        g05\_60.7 & \cellcolor[HTML]{65E14C} & \cellcolor[HTML]{65E14C} \\
        g05\_60.8 & \cellcolor[HTML]{65E14C} & \cellcolor[HTML]{F40B0B} \\
        g05\_60.9 & \cellcolor[HTML]{65E14C} & \cellcolor[HTML]{65E14C} \\
        \hline
        g05\_80.0 & \cellcolor[HTML]{65E14C} & \cellcolor[HTML]{65E14C} \\
        g05\_80.1 & \cellcolor[HTML]{65E14C} & \cellcolor[HTML]{65E14C} \\
        g05\_80.2 & \cellcolor[HTML]{65E14C} & \cellcolor[HTML]{65E14C} \\
        g05\_80.3 & \cellcolor[HTML]{F40B0B} & \cellcolor[HTML]{F40B0B} \\
        g05\_80.4 & \cellcolor[HTML]{65E14C} & \cellcolor[HTML]{65E14C} \\
        g05\_80.5 & \cellcolor[HTML]{F40B0B} & \cellcolor[HTML]{F40B0B} \\
        g05\_80.6 & \cellcolor[HTML]{65E14C} & \cellcolor[HTML]{65E14C} \\
        g05\_80.7 & \cellcolor[HTML]{65E14C} & \cellcolor[HTML]{65E14C} \\
        g05\_80.8 & \cellcolor[HTML]{F40B0B} & \cellcolor[HTML]{F40B0B} \\
        g05\_80.9 & \cellcolor[HTML]{65E14C} & \cellcolor[HTML]{65E14C} \\
        \hline
        g05\_100.0 & \cellcolor[HTML]{65E14C} & \cellcolor[HTML]{65E14C} \\
        g05\_100.1 & \cellcolor[HTML]{65E14C} & \cellcolor[HTML]{65E14C} \\
        g05\_100.2 & \cellcolor[HTML]{65E14C} & \cellcolor[HTML]{65E14C} \\
        g05\_100.3 & \cellcolor[HTML]{F40B0B} & \cellcolor[HTML]{F40B0B} \\
        g05\_100.4 & \cellcolor[HTML]{65E14C} & \cellcolor[HTML]{65E14C} \\
        g05\_100.5 & \cellcolor[HTML]{F40B0B} & \cellcolor[HTML]{F40B0B} \\
        g05\_100.6 & \cellcolor[HTML]{65E14C} & \cellcolor[HTML]{65E14C} \\
        g05\_100.7 & \cellcolor[HTML]{65E14C} & \cellcolor[HTML]{65E14C} \\
        g05\_100.8 & \cellcolor[HTML]{F40B0B} & \cellcolor[HTML]{F40B0B} \\
        g05\_100.9 & \cellcolor[HTML]{65E14C} & \cellcolor[HTML]{65E14C} \\
        \hline
    \end{tabular}
    \label{tab:AdiabaticEasy}
\end{table}

Out of the $30$ problems considered, $23$ are classified as Ising easy and $24$ as adiabatic easy. Notably, every instance that is Ising easy is also adiabatic easy. This suggests that the hybrid continuation method, CAA followed by RA, leads to the ground state of the target Hamiltonian for slightly more problem instances than RA.

\subsection{Numerical simulations with noise}
In order to evaluate the performance of the hybrid CAA method, we perform numerical simulations, integrating Eq.~\ref{eq:tanh_update} with the hybrid CAA schedule. The results are then compared to the RA scheme, which has been shown to perform similarly as other state of the art techniques such as chaotic amplitude control \cite{Paper_Leen_predictingoptimalnoisestrength}. As a performance metric, we use the success rate (SR), which denotes the fraction of runs that end up in the ground state and the time-to-target (TTT), which quantifies how long the IM must run to achieve a $99$\% probability of reaching the ground state. The TTT metric therefore captures both the runtime and the success rate of the algorithm. The formal definition of TTT and details on how it is computed can be found in the Methods section.

The hybrid CAA method combines two annealing stages. first, the parameter $\mathcal{F}$ is gradually increased from zero to one. After this, $\beta$ is increased starting from $\beta_{\text{start}}$. For MaxCut instances, $\beta_{\text{start}}=\beta_{\text{min}}$ and for the Beasley instances, $\beta_{\text{start}}=0.01$. The rate of change in both stages is quantified as follows. In the first annealing stage, $\mathcal{F}$ is initially set to zero and increased to one according to
\begin{equation}
    \label{eq:Adiabatic_annealing_speed}
    \mathcal{F} = v_\mathcal{F} t,
\end{equation}
where $v_\mathcal{F}$ is the adiabatic annealing speed parameter. In the second stage, $\beta$ is incremented at each timestep according to
\begin{equation}
    \label{eq:annealing_speed}
    \beta = \beta_{\text{start}} + v_\beta t,
\end{equation}
where $v_\beta$ is the annealing speed parameter. 

\begin{figure}
    \centering
    \includegraphics[width=\linewidth]{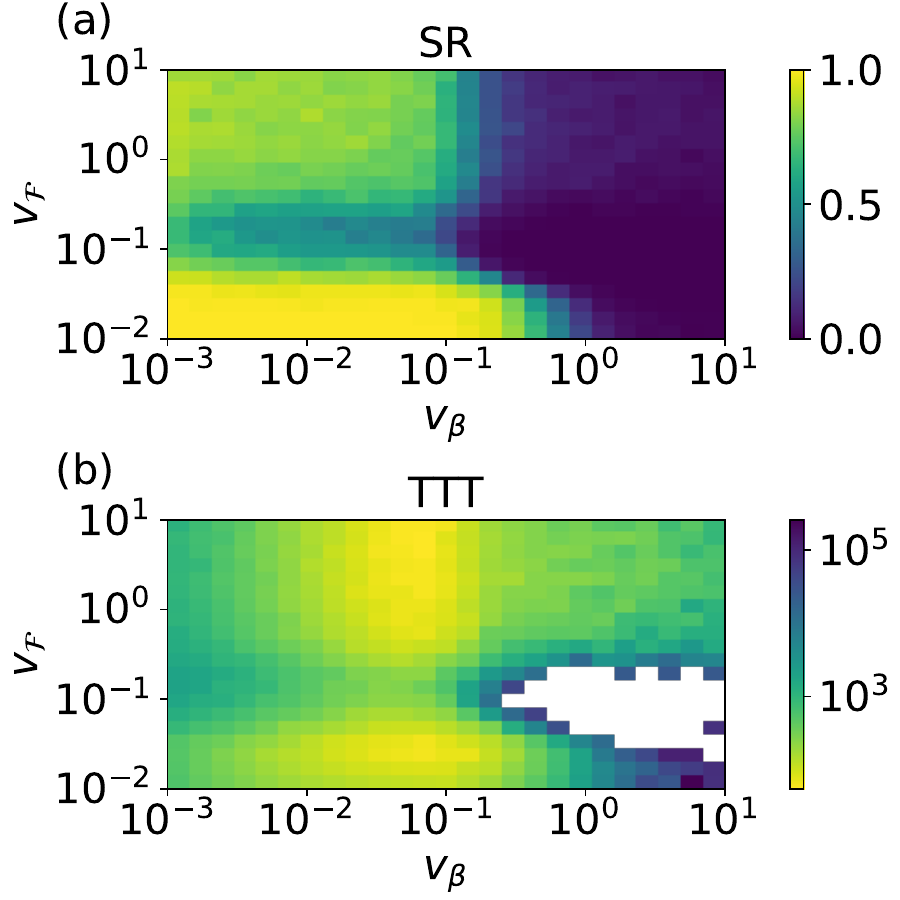}
    \caption{{\bf Parameter scan of the hybrid CAA method for an adiabatic easy problem.} (a) success rate (SR) and (b) time-to-target (TTT) as a function of the annealing speed $v_\beta$ and the adiabatic annealing speed $v_\mathcal{F}$ for problem g05\_60.3, fixed linear gain $\alpha=-3.5$ and noise strength $\gamma = 0.1$. A TTT of infinity is colored white.}
    \label{fig:FullScan_AdiabaticEasy}
\end{figure}

Fig.~\ref{fig:FullScan_AdiabaticEasy} shows the SR and TTT for the example adiabatic easy problem g05\_60.3, plotted as a function of both annealing speeds $v_\beta$ and $v_\mathcal{F}$. In panel (a), the region with the highest SR is at low $v_\beta$ and low $v_\mathcal{F}$. This indicates that when both annealing stages proceed sufficiently slowly, the IM consistently finds the ground state, as is expected for an adiabatic easy problem. Within this region, since the SR is uniformly $100\%$, the lowest TTT is achieved when the annealing is performed as quickly as possible, as illustrated in Fig.~\ref{fig:FullScan_AdiabaticEasy}(b). Remarkably, panel (a) also reveals a second region of high SR in the upper-left corner, corresponding to low $v_\beta$ and high $v_\mathcal{F}$. In this regime, the adiabatic annealing stage is so fast that the IM effectively skips it. Instead, RA is performed directly from the all-spin-up state. As shown in panel (b), this approach can yield a performance comparable to the hybrid method.

\begin{figure}
    \centering
    \includegraphics[width=\linewidth]{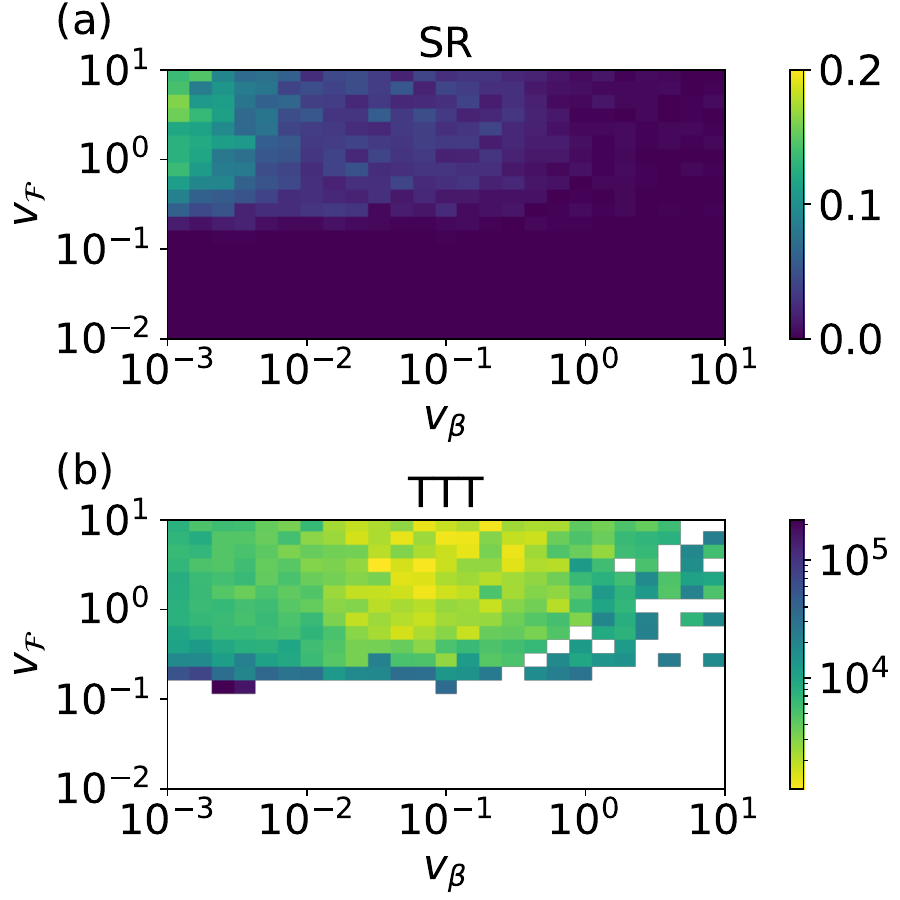}
    \caption{{\bf Parameter scan of the hybrid CAA method for an adiabatic hard problem.}  (a) success rate (SR) and (b) time-to-target (TTT) as a function of the annealing speed $v_\beta$ and the adiabatic annealing speed $v_\mathcal{F}$ for problem g05\_100.3, fixed linear gain $\alpha=-3.5$ and noise strength $\gamma = 0.1$. A TTT of infinity is colored white.}
    \label{fig:FullScan_AdiabaticHard}
\end{figure}

A similar plot is shown for the adiabatic hard problem g05\_100.3 in Fig.~\ref{fig:FullScan_AdiabaticHard}. Contrary to what might be expected based on the continuation results, the SR is not zero everywhere. This is because a run of the IM will neither be perfectly noise-free nor infinitely slow. Deviations from the trajectories obtained by continuation are therefore expected. Here, the SR only drops to zero for lower values of $v_\mathcal{F}$, when the annealing is sufficiently slow for the IM to closely follow the branch predicted by the continuation method. In contrast, performing the annealing more rapidly can modify the energy landscape too quickly for the IM to track its fixed point. Because of this, and aided by noise, the IM can deviate from the branch, resulting in a low, but nonzero SR. 

\begin{figure}
    \centering
    \includegraphics[width=0.99\linewidth]{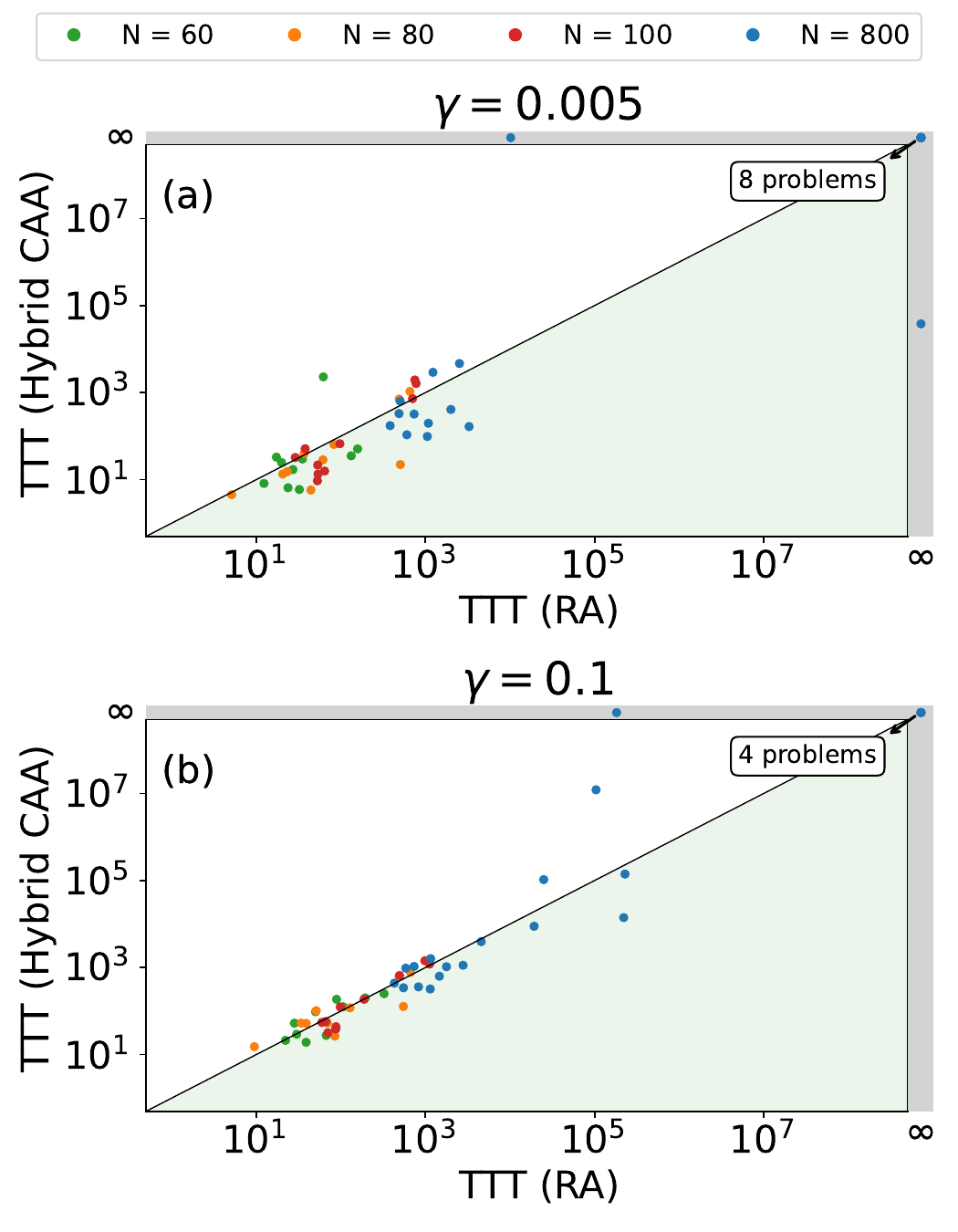}
    \caption{{\bf Performance comparison on MaxCut instances from the BiqMac and GSET libraries.} Comparison of the time-to-target (TTT) of the hybrid classical adiabatic annealing (CAA) method with regular annealing (RA) for a fixed noise strength of (a) $\gamma=0.005$ and (b) $\gamma=0.1$. Each dot represents a problem instance. Its $y$ coordinate represents the TTT when it is solved using the hybrid CAA method and its $x$ coordinate the TTT when solved using RA. So all dots below the diagonal, indicated by the green background, are problems that are solved faster by the hybrid CAA method. Dots in the gray area outside of the plot represent problem instances that cannot be solved by one or by both of the methods (TTT=$\infty$). The TTT values are the lowest value obtained in a grid scan over $\alpha$, the annealing speed $v_\beta$ and $v_\mathcal{F}$.}
    \label{fig:ComparisonAllProblems}
\end{figure}

To gauge the performance of the hybrid CAA method, we perform such a parameter scan for all MaxCut instances. We compare the best TTT obtained in this scan with that for the RA method in Fig.~\ref{fig:ComparisonAllProblems}. We consider the $30$ unweighted BiqMac instances with $50\%$ edge density and problem sizes up to $100$ spins, as well as all $21$ GSET instances with $800$ spins. The GSET problems are more challenging to solve, so a run is considered successful if it reaches at least $99\%$ of the best found energy listed in Ref.~\cite{ma_multiple_2017}. Fig.~\ref{fig:ComparisonAllProblems} (a) compares the performance of the hybrid CAA method with the RA method at a fixed low noise level. Each dot represents a problem instance and its coordinates are the best TTT found via a grid search over the parameters $\alpha$, $v_\beta$ and $v_\mathcal{F}$. Dots located in the shaded border region indicate problems that could not be solved by one or both methods ($\text{TTT}=\infty$). The majority of instances lie below the diagonal (highlighted in green), indicating that, when both methods succeed, the hybrid CAA method generally reaches a solution faster than the RA method. On average, the hybrid CAA method is approximately three times faster. Each method is also able to solve one instance that the other cannot, while eight instances, exclusively larger GSET problems, cannot be solved by either method. Increasing the noise improves the performance of both the hybrid CAA method and the RA method as shown in Fig.~\ref{fig:ComparisonAllProblems}(b). At this higher noise level, only four instances remain unsolved by both methods, compared to eight at lower noise. Now, there is one instance that can be solved with RA, but not with the hybrid CAA. For instances that can be solved by both approaches, the performance advantage of the hybrid CAA method is reduced as the data points lie closer to the diagonal, and the hybrid CAA method is now only about $1.6$ times faster on average. Overall, the two methods perform comparably at higher noise levels, with the hybrid CAA method having the additional overhead of having to optimize the extra parameter $v_\mathcal{F}$.

Next, we investigate whether the presence of external fields influences the relative performance of hybrid CAA and RA. To this end, we consider the Beasley instances of the BiqMac library with problem sizes up to $250$ spins because for these instances, the exact ground state is known \cite{BiqMac}. These instances are quadratic unconstrained binary optimization problems where the objective is to obtain the configuration of binary variables $x_i\in\{0,1\}$ that minimizes
\begin{equation}
    \label{QUBO_Hamiltonian}
    \sum_{ij} Q_{ij}x_ix_j,
\end{equation}
where $Q_{ij}$ is a symmetric matrix implementing the interactions between the variables. This problem can be formulated as an Ising problem by mapping the variables $x_i$ to Ising spins $\sigma_i$ via
\begin{equation}
    x_i = \frac{\sigma_i+1}{2}.
    \label{eq:transfo binary spins to spin}
\end{equation}
This results in an Ising Hamiltonian with $J_{ij}=-\frac{1}{2}\tilde{Q}_{ij}$ and $h_i=-\frac{1}{2}\sum_j Q_{ij}$, where $\tilde{Q}_{ij}$ is $Q_{ij}$, but with all diagonal elements set to zero. A complete derivation of this mapping can be found in Supplementary Note 7 of Ref.~\cite{ExternalFields}. So, although this problem natively does not have any external fields, it does when it is formulated as an Ising problem.

The hybrid CAA procedure is very similar to the procedure used above for the MaxCut problems, with the exception that we now start from an initial Hamiltonian defined by the coupling in Eq.~\ref{eq:InitialCouplingMatrix_AllToAll_Ferromagnetic} and with external fields $h_i=1$. More details can be found in the Methods section. 

In appendix section \ref{subsec:InitialBetaValue_Beasley}, we verified that, similar to MaxCut instances, using lower values of $\beta_\text{start}$ leads to a better performance for most of the Beasley instances. As problems with external fields do not have a minimum $\beta$ value required to avoid the collapse of the spin amplitudes to zero, we use the arbitrary low value of $\beta_\text{start}=0.01$.



We perform the same grid search over the parameters $\alpha$, $v_\beta$, and $v_\mathcal{F}$ for all Beasley problems, and report the best time-to-target (TTT) obtained. Fig.~\ref{fig:ComparisonBeasleyProblems}(a) shows a comparison of this TTT of the hybrid CAA method and of the RA method. Most instances lie very close to the diagonal, indicating they are solved in a similar time for both methods. However, there are some instances that are solved up to two orders of magnitude faster with the hybrid CAA method than with RA. This indicates that, similar to the MaxCut instances, the hybrid CAA method slightly outperforms RA.

However, when using IMs that employ continuous variables to implement the Ising spins, an imbalance between the spin coupling term $\sim J_{ij}s_j$ and the external fields $\sim h_i$ can occur when the spin amplitudes are small, causing the IM to effectively ignore the spin coupling. Previous work showed that replacing the spin amplitude by its sign in the spin coupling term resolves this imbalance, drastically improving the performance of the IM when using RA \cite{ExternalFields}. We refer to this as the spin sign method, in which case the evolution of the spin amplitudes are modeled according to
\begin{multline}
    \label{eq:tanh_spinsign_update}
    \frac{ds_i}{dt} = -s_i\\  + \tanh\left(\alpha s_i + \beta\left(\sum_j^N J_{ij}\sigma_j + h_i\right)+\gamma\xi_i(t)\right).
\end{multline}
As employing the spin sign method improves the performance of RA, Fig.~\ref{fig:ComparisonBeasleyProblems}(a) actually does not compare the hybrid CAA method to the state-of-the-art performance of RA. That being said, the spin sign method can also be applied to the hybrid CAA method. Therefore, Fig.~\ref{fig:ComparisonBeasleyProblems}(b) shows a similar performance comparison when both the hybrid CAA method and RA employ the spin sign method. All dots are shifted towards the lower left compared to Fig.~\ref{fig:ComparisonBeasleyProblems}(a), indicating that for both methods, most instances are solved faster when using the spin sign method. In this setting, the performance advantage of hybrid CAA largely disappears. The hybrid CAA method and RA show very similar behavior, with each slightly outperforming the other on different instances. We therefore conclude that, when augmented with the spin sign method, the hybrid CAA approach and RA achieve comparable performance, similarly to the MaxCut instances.

\begin{figure}
    \centering
    \includegraphics[width=\linewidth]{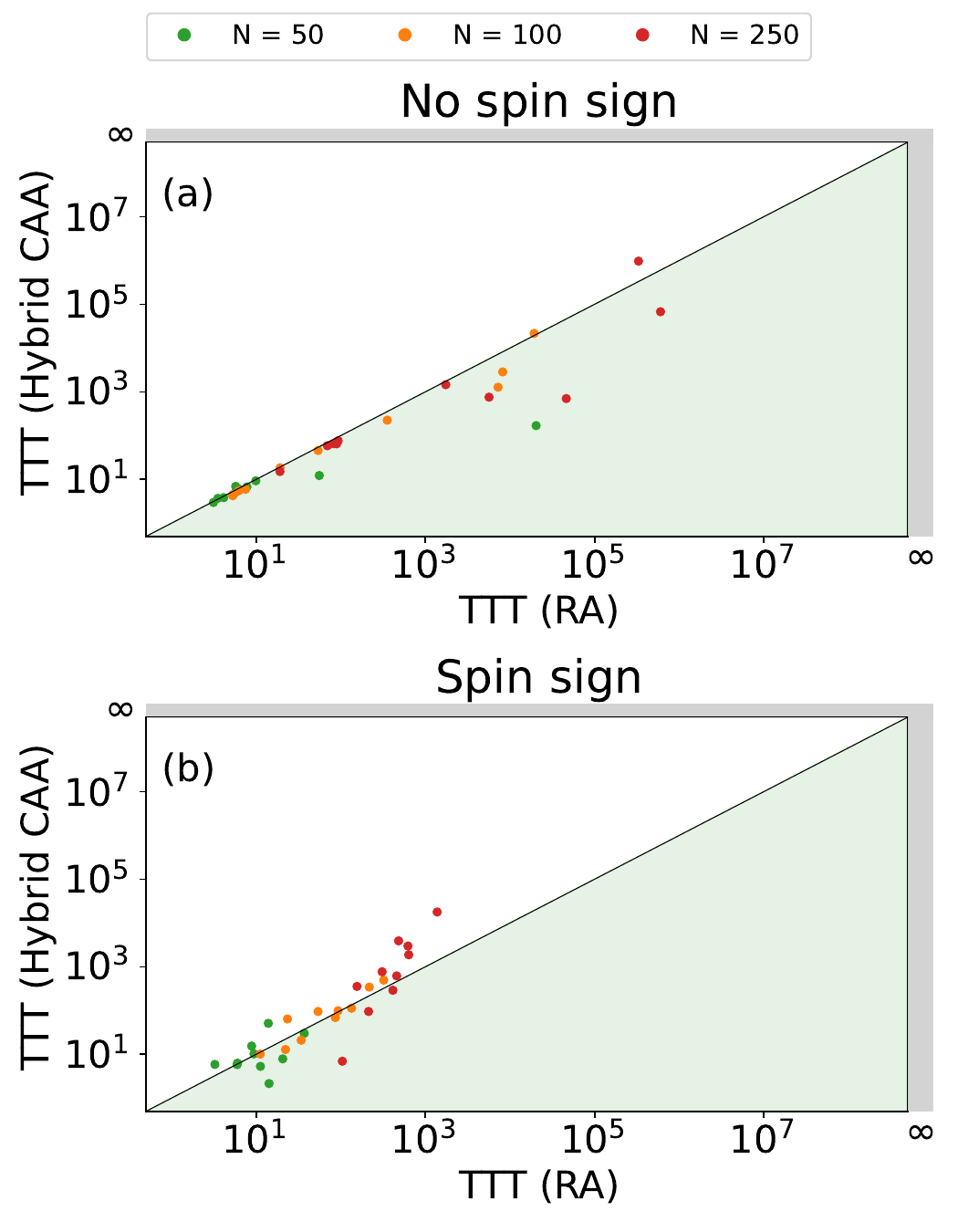}
    \caption{{\bf Performance comparison on Beasley instances.} Comparison of the time-to-target (TTT) of the hybrid CAA with regular annealing (RA) when employing the spin sign method (b) or (a) not. Each dot represents a problem instance. Its $y$ coordinate represents the TTT when it is solved using the hybrid CAA method and its $x$ coordinate the TTT when solved using RA. Dots below the diagonal, indicated by the green background, are problems that are solved faster by the hybrid CAA method. Dots in the gray area outside of the plot represent problem instances that cannot be solved by one of the methods or by both (TTT=$\infty$). The TTT values are the lowest value obtained in a grid scan over $\alpha$, the annealing speed $v_\beta$ and $v_\mathcal{F}$, for a fixed noise strength of $\gamma=0.1$.}
    \label{fig:ComparisonBeasleyProblems}
\end{figure}

\section{Discussion}
When performing CAA, the IM is initialized in the ground state of an initial Hamiltonian whose ground state is easily obtained. This Hamiltonian is then gradually deformed into the target Hamiltonian representing the optimization problem. To better understand the behavior of this process, we performed a continuation analysis, which reveals intrinsic limitations of CAA. In particular, a typical annealing trajectory encounters SNs that can cause the system to leave the desired solution branch. Avoiding such disruptive SNs requires reducing the coupling strength $\beta$. However, lowering $\beta$ introduces a competing issue as the ground state configuration of the target Hamiltonian may no longer correspond to a stable fixed point. Consequently, there is no value of $\beta$ for which CAA directly leads to the ground state fixed point without encountering any SNs.
 
To overcome this issue, we propose the hybrid CAA method, which consists of two distinct stages: the CAA stage and the Regular Annealing (RA) stage. In the first stage, the IM is initialized in the ground state of a fully ferromagnetic Hamiltonian and evolved toward the target Hamiltonian using a coupling strength $\beta$ that is chosen as small as possible to avoid SNs. For MaxCut instances, this choice is constrained by the need to prevent collapse of the spin amplitudes to zero, which defines a lower bound on $\beta_\text{start}$. Because this value of $\beta$ is typically too small for the ground state of the target Hamiltonian to correspond to a stable fixed point, the CAA stage is followed by the RA stage. In this second stage, $\beta$ is gradually increased, allowing the system to converge toward a configuration corresponding to a minimum of the binary Hamiltonian. We have shown that our hybrid CAA method renders many of the MaxCut instances of the BiqMac library adiabatic easy, a feat standard CAA could not achieve.

The hybrid CAA method is compatible with any hardware implementation of analog Ising machines. However, its initialization step, starting from a Hamiltonian in which all spins are ferromagnetically coupled, requires all-to-all connectivity, which may not be feasible in certain physical setups. In practice, coupling each spin to as many others as the hardware allows may still yield effective results. However, verifying this would require a detailed study focused on a specific hardware architecture, which lies beyond the scope of this work. Furthermore, the hybrid CAA method requires the coupling matrix and external fields to be altered every timestep, which may be challenging or time consuming for some architectures.

We demonstrated that the hybrid CAA method outperforms RA in general, but only by a slight margin. It achieves an average speedup of $1.6$ for MaxCut and $8$ for Beasley instances for $\gamma=0.1$ compared to RA. However, these performance improvements come at the cost of an additional hyperparameter ($v_\mathcal{F}$) to optimize. Furthermore, when applying the spin sign method to both the hybrid CAA method and RA, the performance gap for the Beasley instances largely disappears and the two methods perform very similarly. Overall, we believe that the hardware challenges, combined with adding the extra parameter $v_\mathcal{F}$ to optimize, offsets the potential benefits of using this method.

\section{Methods}
\label{sec:Methods}

\subsection{Continuation}
The continuation of the CAA starts from the ground state of the initial ferromagnetic Hamiltonian with a coupling matrix given by Eq.~(\ref{eq:InitialCouplingMatrix_AllToAll_Ferromagnetic}). The fixed point that corresponds to the binary ground state of this system is the state in which all spin amplitudes have the same sign and amplitude $c$. The value of $c$ is such that the right-hand side of Eq.~(\ref{eq:tanh_update}) is zero and is determined via Newtons method. This amplitude depends on both system parameters $\alpha$ and $\beta$. 

The initial coupling strength $\beta_\text{start}$ is always chosen to be $\beta_{\text{min}}$, defined in Eq.~(\ref{eq:beta_min}). The continuation was performed for all problems for the following values of the linear gain parameter $\alpha$: $0.999$, $0.9$, $0.7$, $0.5$, $0.$, $-1$, $-2$, $-5$, $-10$, $-20$ and $-50$. The evolution equations of Eq.~(\ref{eq:tanh_update}) with the coupling matrix given by Eq.~(\ref{eq:CouplingMatrixCAA}) are submitted to the auto-07p software. $\mathcal{F}$ is treated as a third system parameter and is increased from zero to one. The fixed point obtained at $\mathcal{F}=1$ is then used as the initial state for the RA. Now, $\beta$ is increased from $\beta_{\text{min}}$ while the other parameters are kept constant. The final value of $\beta$ depends on the value of $\alpha$, but is large enough to ensure that all spins are saturated at $-1$, $0$ or $1$ at the end of the continuation.

\subsection{Integration of the evolution equations}
\begin{table*}[!ht]
\caption{\textbf{Parameter sweeps.} The ranges of the parameters used to determine their optimal values.}
\centering
\renewcommand{\arraystretch}{1.2}
\begin{tabular}{|p{0.13\linewidth}|p{0.12\linewidth}|p{0.10\linewidth}|p{0.10\linewidth}|p{0.11\linewidth}|p{0.1\linewidth}|p{0.1\linewidth}|}
\hline
\textbf{Annealing schedule} & \textbf{Parameter name} & \textbf{Lowest value} & \textbf{Highest value} & \textbf{Spacing type} & \textbf{Number of points BiqMac} & \textbf{Number of points GSET} \\
\hline\hline
\multirow{3}{*}{CAA} 
    & $\alpha$   & $-3.5$       & $0.7$         & Linear           & $7$ & $7$\\
    & $v_\beta$ & $10^{-3}$ & $10^{1}$ & Log (base 10) & $10$ & $5$ \\
    & $v_\mathcal{F}$  & $10^{1}$   & $10^{4}$   & Log (base 10)    & $10$ & $5$ \\
\hline
\multirow{2}{*}{RA} 
    & $\alpha$ & $-3.5$       & $0.7$         & Linear           & $7$ & $7$\\
    & $v_\beta$ & $10^{-3}$ & $10^{1}$ & Log (base 10) & $10$ & $5$ \\
\hline
\end{tabular}
\label{tab:OptimalParameters}
\end{table*}

To simulate the CAA on the IM, we use the same initial state and value for the coupling strength $\beta$ as in the continuation procedure. The system evolves according to the update equations in Eq.~(\ref{eq:tanh_update}), which are integrated using an Euler scheme:
\begin{multline}
    \label{eq:Euler_update}
    s^{t+1}_i = s_i^t + \\\Delta t \left(-s^t_i + \tanh\left(\alpha s_i^t + \beta\sum_j^N J_{ij}s_i^t\right) \right) + \sqrt{\Delta t}\gamma \xi_i^t,
\end{multline}
where $s_i^t$ is the amplitude of spin $i$ at timestep $t$, $\Delta t=0.01$ is the Euler timestep, $\gamma$ is the noise strength and $\xi_i^t$ is the noise, drawn from a Gaussian distribution with zero mean and a standard deviation of one. The coupling matrix at timestep $t$ is given by
\begin{equation}
    \label{eq:UpdateCouplingMatrix}
    J_{ij} = (1-\mathcal{F}^t)J^{ \text{Initial}}_{ij}+\mathcal{F}^tJ^{ \text{Target}}_{ij},
\end{equation}
where $\mathcal{F}^t$ denotes the $\mathcal{F}$ value at timestep $t$ of the adiabatic annealing stage, given by
\begin{equation}
    \label{eq:Update_F}
    \mathcal{F}^{t+1} = \mathcal{F}^t + v_\mathcal{F}\Delta t,
\end{equation}
with $\mathcal{F}^0 = 0$. For the Beasley instances, the external fields at timestep $t$ are given by
\begin{equation}
    \label{eq:UpdateExternalFields}
    h_{i} = (1-\mathcal{F}^t)h^{ \text{Initial}}_{i}+\mathcal{F}^th^{ \text{Target}}_{i},
\end{equation}
where $h^{\text{Target}}_{i}$ are the external fields defined by the problem instance and $h^{ \text{Initial}}_{i} = 1$ $\forall i$.


During the RA stage, $\mathcal{F}$ is fixed at $1$ and $\beta$ is increased every time step according to
\begin{equation}
    \label{eq:RA_beta}
    \beta^{t+1}=\beta^t + v_\beta \Delta t,
\end{equation}
where $v_\beta$ is the annealing speed parameter. RA is performed until either the ground state solution is found or $\beta$ exceeds $10$.

For each problem instance, a parameter sweep is performed over $\alpha$, $v_\mathcal{F}$, and $v_\beta$, with the specific ranges summarized in Table \ref{tab:OptimalParameters}. For every combination of parameters, the IM is initialized and run $500$ times. If the ground state energy is reached during a run, it is terminated early, and the simulated time taken to reach that energy is recorded. The success rate $P$ is then computed as the fraction of runs that successfully reached the ground state, or $99\%$ of the ground state in case of the GSET instances. To evaluate performance, the time-to-target (TTT) is computed as a function of the runtime $T$, ranging from zero up to the maximum observed time required to reach the ground state. The runtime is the time of performing both annealing stages combined. Instead of rerunning the IM for each value of $T$, the recorded completion times from the $500$ runs are used to determine how many runs would have finished before each value of the runtime $T$. This yields an estimation of the success rate $P(T)$ for each $T$, from which the corresponding TTT is calculated using:
\begin{align}
\label{eq:TTT}
    \text{TTT}(T) = \begin{cases}
        T&P(T)>0.99\\
        T\frac{\log(0.01)}{\log(1-P(T))} &0<P(T)\leqslant 0.99\\
        \infty & P(T)=0        
    \end{cases}.
\end{align}
The TTT value reported is the minimum of $\text{TTT}(T)$ over all values of $T$.

\section{Data availability}
The authors declare that all relevant data are included in the manuscript. Additional data are available from the corresponding author upon reasonable request.

\section{Author contributions}
J.L. performed the simulations and wrote the manuscript. G.V. and G.V.d.S. supervised the project. All authors discussed the results and reviewed the manuscript.

\section{Additional information}
{\bf Competing interests:} 
All authors declare no competing interests.\\
{\bf Acknowledgements:} The authors would like to thank Toon Sevenants for the insightful discussions. \\This research was funded by the Research Foundation Flanders (FWO) under grants G028618N, G029519N, G0A6L25N and G006020N. Additional funding was provided by the EOS project "Photonic Ising Machines". This project (EOS number 40007536) has received funding from the FWO and F.R.S.-FNRS under the Excellence of Science (EOS) programme. The computational resources and services used in this work were provided by the VSC (Flemish Supercomputer Center), funded by the Research Foundation Flanders (FWO) and the Flemish Government.

\clearpage

\printbibliography

@article{Poor_mans_CIM,
  title={A poor man’s coherent Ising machine based on opto-electronic feedback systems for solving optimization problems},
  author={B{\"o}hm, Fabian and Verschaffelt, Guy and Van der Sande, Guy},
  journal={Nature communications},
  volume={10},
  number={1},
  pages={3538},
  year={2019},
  publisher={Nature Publishing Group UK London}
}

@article{Order_of_magnitude,
  title={Order-of-magnitude differences in computational performance of analog Ising machines induced by the choice of nonlinearity},
  author={B{\"o}hm, Fabian and Vaerenbergh, Thomas Van and Verschaffelt, Guy and Van der Sande, Guy},
  journal={Communications Physics},
  volume={4},
  number={1},
  pages={149},
  year={2021},
  publisher={Nature Publishing Group UK London}
}

@article{Berloff,
  title={Computational complexity continuum within Ising formulation of NP problems},
  author={Kalinin, Kirill P and Berloff, Natalia G},
  journal={Communications Physics},
  volume={5},
  number={1},
  pages={20},
  year={2022},
  publisher={Nature Publishing Group UK London}
}

@article{Ising_formulations_of_many_NP_problems,
  title={Ising formulations of many NP problems},
  author={Lucas, Andrew},
  journal={Frontiers in physics},
  volume={2},
  pages={74887},
  year={2014},
  publisher={Frontiers}
}

@article{2000nodes,
  title={A coherent Ising machine for 2000-node optimization problems},
  author={Inagaki, Takahiro and Haribara, Yoshitaka and Igarashi, Koji and Sonobe, Tomohiro and Tamate, Shuhei and Honjo, Toshimori and Marandi, Alireza and McMahon, Peter L and Umeki, Takeshi and Enbutsu, Koji and others},
  journal={Science},
  volume={354},
  number={6312},
  pages={603--606},
  year={2016},
  publisher={American Association for the Advancement of Science}
}

@article{Quantum_annealing_DWave,
  title={Quantum annealing with manufactured spins},
  author={Johnson, Mark W and Amin, Mohammad HS and Gildert, Suzanne and Lanting, Trevor and Hamze, Firas and Dickson, Neil and Harris, Richard and Berkley, Andrew J and Johansson, Jan and Bunyk, Paul and others},
  journal={Nature},
  volume={473},
  number={7346},
  pages={194--198},
  year={2011},
  publisher={Nature Publishing Group UK London}
}

@MISC{AUTO,
author = "Eusebius J. Doedel and Thomas Fairgrieve and Björn Sandstede and Alan R. Champneys and Yuri Kuznetsov and Xianjun Wang",
title = "AUTO-07p: Continuation and bifurcation
software for ordinary differential equations",
year = "2007",
 }

@article{Conti,
  title={Large-scale photonic Ising machine by spatial light modulation},
  author={Pierangeli, D and Marcucci, G and Conti, C},
  journal={Physical review letters},
  volume={122},
  number={21},
  pages={213902},
  year={2019},
  publisher={APS}
}

@article{paper_Ganguli,
  title = {Geometric Landscape Annealing as an Optimization Principle Underlying the Coherent Ising Machine},
  author = {Yamamura, Atsushi and Mabuchi, Hideo and Ganguli, Surya},
  journal = {Phys. Rev. X},
  volume = {14},
  issue = {3},
  pages = {031054},
  year = {2024},
  doi = {10.1103/PhysRevX.14.031054}
}

@article{Traffic_Flow,
  title={Traffic flow optimization using a quantum annealer},
  author={Neukart, Florian and Compostella, Gabriele and Seidel, Christian and Von Dollen, David and Yarkoni, Sheir and Parney, Bob},
  journal={Frontiers in ICT},
  volume={4},
  pages={29},
  year={2017},
  publisher={Frontiers Media SA}
}

@article{Fincance,
  title={Quantum computing for finance: Overview and prospects},
  author={Or{\'u}s, Rom{\'a}n and Mugel, Samuel and Lizaso, Enrique},
  journal={Reviews in Physics},
  volume={4},
  pages={100028},
  year={2019},
  publisher={Elsevier}
}

@article{Job_Scheduling,
  title={A case study in programming a quantum annealer for hard operational planning problems},
  author={Rieffel, Eleanor G and Venturelli, Davide and O’Gorman, Bryan and Do, Minh B and Prystay, Elicia M and Smelyanskiy, Vadim N},
  journal={Quantum Information Processing},
  volume={14},
  pages={1--36},
  year={2015},
  publisher={Springer}
}

@article{Protein_Folding,
  title={Construction of model Hamiltonians for adiabatic quantum computation and its application to finding low-energy conformations of lattice protein models},
  author={Perdomo, Alejandro and Truncik, Colin and Tubert-Brohman, Ivan and Rose, Geordie and Aspuru-Guzik, Al{\'a}n},
  journal={Physical Review A},
  volume={78},
  number={1},
  pages={012320},
  year={2008},
  publisher={APS}
}

@online{BiqMac,
  author = {Angelika Wiegele},
  title = {Biq Mac Library - A collection of Max-Cut and quadratic 0-1 programming instances of medium size},
  year = 2007,
  url = {https://biqmac.aau.at/biqmaclib.pdf}
}

@online{GSET,
    author = {Y. Ye},
    title={GSET},
    url = {https://web.stanford.edu/~yyye/yyye/Gset/},
    note={Accessed: 2025-05-21}
}

@article{100000SpinsCIM,
  title={100,000-spin coherent Ising machine},
  author={Honjo, Toshimori and Sonobe, Tomohiro and Inaba, Kensuke and Inagaki, Takahiro and Ikuta, Takuya and Yamada, Yasuhiro and Kazama, Takushi and Enbutsu, Koji and Umeki, Takeshi and Kasahara, Ryoichi and others},
  journal={Science advances},
  volume={7},
  number={40},
  pages={eabh0952},
  year={2021},
  publisher={American Association for the Advancement of Science}
}

@article{OverviewMcMahon,
  title={Ising machines as hardware solvers of combinatorial optimization problems},
  author={Mohseni, Naeimeh and McMahon, Peter L and Byrnes, Tim},
  journal={Nature Reviews Physics},
  volume={4},
  number={6},
  pages={363--379},
  year={2022},
  publisher={Nature Publishing Group UK London}
}

@article{PolaritonCondensates,
  title={Realizing the classical XY Hamiltonian in polariton simulators},
  author={Berloff, Natalia G and Silva, Matteo and Kalinin, Kirill and Askitopoulos, Alexis and T{\"o}pfer, Julian D and Cilibrizzi, Pasquale and Langbein, Wolfgang and Lagoudakis, Pavlos G},
  journal={Nature materials},
  volume={16},
  number={11},
  pages={1120--1126},
  year={2017},
  publisher={Nature Publishing Group UK London}
}

@article{PolaritonCondensates_2,
  title={Global optimization of spin Hamiltonians with gain-dissipative systems},
  author={Kalinin, Kirill P and Berloff, Natalia G},
  journal={Scientific reports},
  volume={8},
  number={1},
  pages={17791},
  year={2018},
  publisher={Nature Publishing Group UK London}
}

@article{DigitalAnnealer,
author = {Aramon, Maliheh and Rosenberg, Gili and Valiante, Elisabetta and Miyazawa, Toshiyuki and Tamura, Hirotaka and Katzgraber, Helmut},
year = {2019},
month = {04},
pages = {48},
title = {Physics-Inspired Optimization for Quadratic Unconstrained Problems Using a Digital Annealer},
volume = {7},
journal = {Frontiers in Physics},
doi = {10.3389/fphy.2019.00048}
}

@article{Ising_machine_based_on_networks_of_subharmonic_electrical_resonators,
  title = {An Ising machine based on networks of subharmonic electrical resonators},
  author = {English, L. Q. and Zampetaki, A. V. and Kalinin, K. P. and Berloff, N. G. and Kevrekidis, P. G.},
  journal = {Communications Physics},
  volume = {5},
  issue = {1},
  year = {2022},
}

@article{Mean-field_CIM_with_artificial_ZeemanTerms,
  title = {Mean-field coherent Ising machines with artificial Zeeman terms},
  author = {Gunathilaka, Mastiyage Don Sudeera Hasaranga and Inui, Yoshitaka and Kako, Satoshi and Yamamoto, Yoshihisa and Aonishi, Toru},
  journal = {J. Appl. Phys},
  volume = {134},
  issue = {23},
  year = {2023},
}

@article{PaperJacob_UsingContinuationMethods,
  title = {Using continuation methods to analyse the difficulty of problems solved by Ising machines},
  author = {Lamers, Jacob and Verschaffelt, Guy and Van der Sande, Guy},
  journal = {Communications Physics},
  volume = {7},
  issue = {1},
  year = {2024},
}

@article{SupplyChainLogisticsWithAnnealing,
  title={Supply chain logistics with quantum and classical annealing algorithms},
  author={Weinberg, Sean J. and Sanches, Fabio and Ide, Takanori and Kamiya, Kazumitzu and Correll, Randall},
  journal={Scientific Reports},
  volume={13},
  issue={1},
  DOI={10.1038/s41598-023-31765-8},
  year={2023}
}

@article{ComputationalBiology,
  title={Quantum annealing versus classical machine learning applied to a simplified computational biology problem},
  author={Li, Richard Y. and Di Felice, Rosa and Rohs, Remo and Lidar, Daniel A.},
  journal={npj Quantum Information},
  volume={4},
  issue={1},
  DOI={10.1038/s41534-018-0060-8},
  year={2018}
}

@article{Probabilistic_computing_with_pbits,
  title={Probabilistic computing with p-bits},
  author={Kaiser, Jan and Datta, Supriyo},
  journal={Applied Physics Letters},
  volume={119},
  issue={15},
  DOI={https://doi.org/10.1063/5.0067927},
  year={2021}
}

@article{Stochastic_pbits_for_invertible_logic,
  title = {Stochastic $p$-Bits for Invertible Logic},
  author = {Camsari, Kerem Yunus and Faria, Rafatul and Sutton, Brian M. and Datta, Supriyo},
  journal = {Phys. Rev. X},
  volume = {7},
  issue = {3},
  pages = {031014},
  year = {2017},
  publisher = {American Physical Society},
  doi = {10.1103/PhysRevX.7.031014},
  
}

@article{jiang2023efficient,
  title={Efficient combinatorial optimization by quantum-inspired parallel annealing in analogue memristor crossbar},
  author={Jiang, Mingrui and Shan, Keyi and He, Chengping and Li, Can},
  journal={Nature communications},
  volume={14},
  number={1},
  pages={5927},
  year={2023},
  publisher={Nature Publishing Group UK London}
}

@article{inspiration_idea_thomas,
  title={High-performance combinatorial optimization based on classical mechanics},
  author={Goto, Hayato and Endo, Kotaro and Suzuki, Masaru and Sakai, Yoshisato and Kanao, Taro and Hamakawa, Yohei and Hidaka, Ryo and Yamasaki, Masaya and Tatsumura, Kosuke},
  journal={Science Advances},
  volume={7},
  number={6},
  pages={eabe7953},
  year={2021},
  publisher={American Association for the Advancement of Science}
}

@article{Integrated_pbits,
author = {Duffee, Christian and Athas, Jordan and Shao, Yixin and Melendez, Noraica and Raimondo, Eleonora and Katine, Jordan and Çamsarı, Kerem and Finocchio, Giovanni and Amiri, Pedram},
year = {2024},
title = {Integrated probabilistic computer using voltage-controlled magnetic tunnel junctions as its entropy source},
journal={ArXiv},
doi = {10.48550/arXiv.2412.08017}
}

@article{ExternalFields,
author = {De Prins, Robbe and Lamers, Jacob and Bienstman, Peter and Verschaffelt, Guy and Van der Sande, Guy and Van Vaerenberg, Thomas},
year = {2025},
title = {How to Incorporate External Fields in Analog Ising Machines},
journal={ArXiv},
doi = {10.48550/arXiv.2505.08796}
}

@article{QuantumAdiabaticAnnealingOnRandomProblems,
author = {Edward Farhi  and Jeffrey Goldstone  and Sam Gutmann  and Joshua Lapan  and Andrew Lundgren  and Daniel Preda },
title = {A Quantum Adiabatic Evolution Algorithm Applied to Random Instances of an NP-Complete Problem},
journal = {Science},
volume = {292},
number = {5516},
pages = {472-475},
year = {2001},
doi = {10.1126/science.1057726}}

@article{QuantumAdiabaticAnnealingOverview,
author = {E Santoro, Giuseppe and Tosatti, Erio},
title = {Optimization using quantum mechanics: quantum annealing through adiabatic evolution},
journal = {Journal of Physics A: Mathematical and General},
volume = {39},
number = {36},
year = {2006},
doi = {10.1088/0305-4470/39/36/R01}}

@article{Paper_Leen_predictingoptimalnoisestrength,
  title = {Predicting the optimal noise strength for solving optimization problems with analog Ising machines},
  author = {Mys, Leen and Verschaffelt, Guy and Van der Sande, Guy},
  journal = {Phys. Rev. Appl.},
  volume = {25},
  issue = {3},
  pages = {034068},
  numpages = {20},
  year = {2026},
  month = {Mar},
  publisher = {American Physical Society},
  doi = {10.1103/czsz-z8ng}
}

@article{Pierangeli_AdiabaticAnnealing,
author = {Davide Pierangeli and Giulia Marcucci and Claudio Conti},
journal = {Optica},
number = {11},
pages = {1535--1543},
publisher = {Optica Publishing Group},
title = {Adiabatic evolution on a spatial-photonic Ising machine},
volume = {7},
year = {2020},
doi = {10.1364/OPTICA.398000},
}

@misc{CAA_Thomas,
    author = {S. Kumar and T. Van Vaerenbergh and J. P. Strachan},
    title = {Classical Adiabatic Annealing in Memristor Hopfield Neural Networks for Combinatorial Optimization},
    note={2020 International Conference on Rebooting Computing (ICRC), Atlanta, GA, USA},
    year = {2020},
    pages = {76-79},
    doi = {10.1109/ICRC2020.2020.00016}
}

@article{ma_multiple_2017,
	title = {A multiple search operator heuristic for the max-k-cut problem},
    author={Ma, Fuda and Hao, Jin-Kao},
  journal={Annals of Operations Research},
  volume={248},
  number={1},
  pages={365--403},
  year={2017},
  publisher={Springer}
}

@article{ApplicationsInPowerSystemOperations,
author = {Kirihara, Kenta and Imai, Hidetaka and Kuroda, Eisuke and Yamazaki, Jun and Masaki-Kato, Akiko},
year = {2023},
month = {06},
pages = {68004 - 68017},
title = {Exploring Potential Applications of Ising Machines for Power System Operations},
volume = {11},
journal = {IEEE Access},
doi = {10.1109/ACCESS.2023.3289720}
}

@article{200GOPS_PhotonicIsingMachine,
author = {Al-Kayed, Nayem and St-Arnault, Charles and Morison, Hugh and Aadhi, A. and Huang, Chaoran and Tait, Alexander N. and Plant, David V. and Shastri, Bhavin J.},
journal = {Nature},
pages = {576-584},
title = {Programmable 200 GOPS Hopfield-inspired photonic Ising machine},
volume = {648},
year = {2025},
doi = {10.1364/OPTICA.398000},
}

\clearpage

\section{Appendix}
\subsection{Initial Hamiltonian}

A natural choice for the initial Hamiltonian is one whose ground state is easily found. A class of such problems is known as spectral easy problems \cite{PaperJacob_UsingContinuationMethods}, where the principal eigenvector of the coupling matrix points to the ground state \cite{Berloff}. In this work, we explore four different spectral easy initial coupling matrices.

The first approach uses the connectivity of the target coupling matrix but imposes ferromagnetic interactions between spins. This results in an initial coupling matrix that is essentially the absolute value of the target matrix. However, this method fails for unweighted antiferromagnetic target matrices, such as those of MaxCut problems. In these cases, the coupling matrix becomes zero when $\mathcal{F} = 0.5$, requiring $\alpha > 1$ to stay above the the critical $\beta$ value. This restricts the method to parameter regions with poor performance, making it unsuitable.

The second option involves a periodic $L \times L$ grid where each spin is antiferromagnetically coupled to its four neighbors. However, as the initial coupling matrix should have the same number of spins as the target coupling matrix, this would limit the target coupling matrices to the set for which $N$ can be written as $L^2$. Because of this limitation, we discarded this approach.

In a third option, spins are arranged on a circle and each spin is ferromagnetically coupled to its two neighbors and the spin on the opposite site of the circle. For this initial condition, there is an issue, illustrated in Fig.~\ref{fig:MobiusGraph}. Fig.~\ref{fig:MobiusGraph}(a) shows the continuation of all spin amplitudes for an exemplary problem. Fig.~\ref{fig:MobiusGraph} shows the evolution of spin amplitude $6$, used as a proxy for the entire fixed point, as a function of $\mathcal{F}$. At $\mathcal{F}=0$, the continuation starts from the ground state of the ferromagnetic Möbius graph and all spin amplitudes have the same magnitude and sign ($0.48$). As $\mathcal{F}$ increases, this fixed point disappears via a saddle-node bifurcation (SN 1), and the resulting unstable branch loops back to lower $\mathcal{F}$ values, reaching a second bifurcation (SN 2). During this evolution, many spin amplitudes change sign, leading to a qualitatively different fixed point. Crucially, this unstable branch persists across all problems and cannot be removed for any value of $\alpha$, rendering all problems adiabatic hard. Consequently, this option was also rejected.

Among the options considered, the all-to-all ferromagnetic coupling matrix defined in Eq.~\ref{eq:InitialCouplingMatrix_AllToAll_Ferromagnetic} proved to be the most effective and robust choice for the initial coupling matrix.

\begin{figure*}[t]
    \centering
    \includegraphics[width=0.9\linewidth]{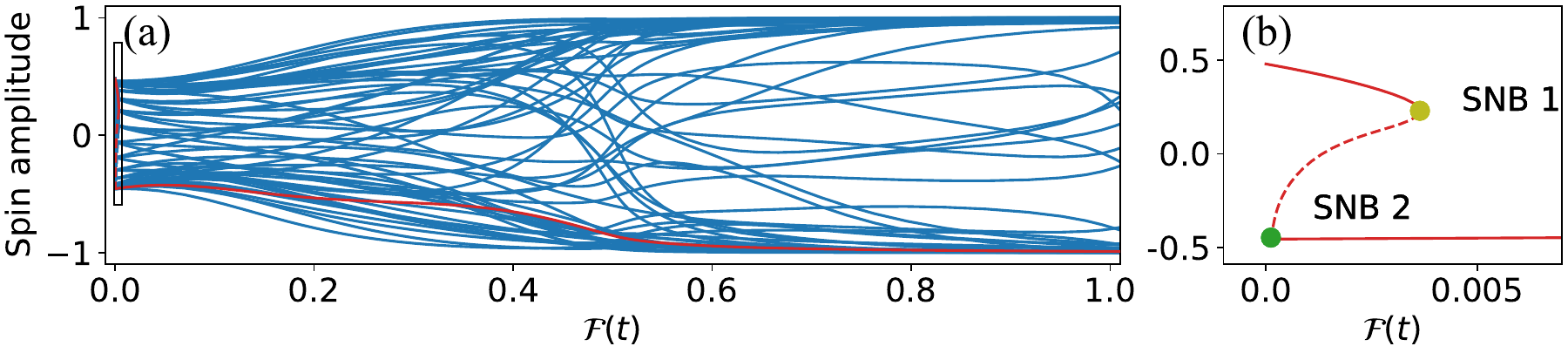}
    \caption{{\bf Continuation of CAA starting from a ferromagnetic Möbius graph.} Evolution as $\mathcal{F}$ varies from zero to one of (a) the spin amplitudes of problem g05\_60.0 for $\alpha=-0.5$ and $\beta=\beta_\text{min}$, (b) spin amplitude 6 in the black rectangle of Fig. (a). Solid (dashed) lines are used when the spin amplitude is part of a stable (unstable) state. SN (green and yellow dots) indicate saddle-node bifurcations}
    \label{fig:MobiusGraph}
\end{figure*}

\subsection{Initial $\beta$ value for the Beasley instances}
\label{subsec:InitialBetaValue_Beasley}

\begin{figure}
    \centering
    \includegraphics[width=\linewidth]{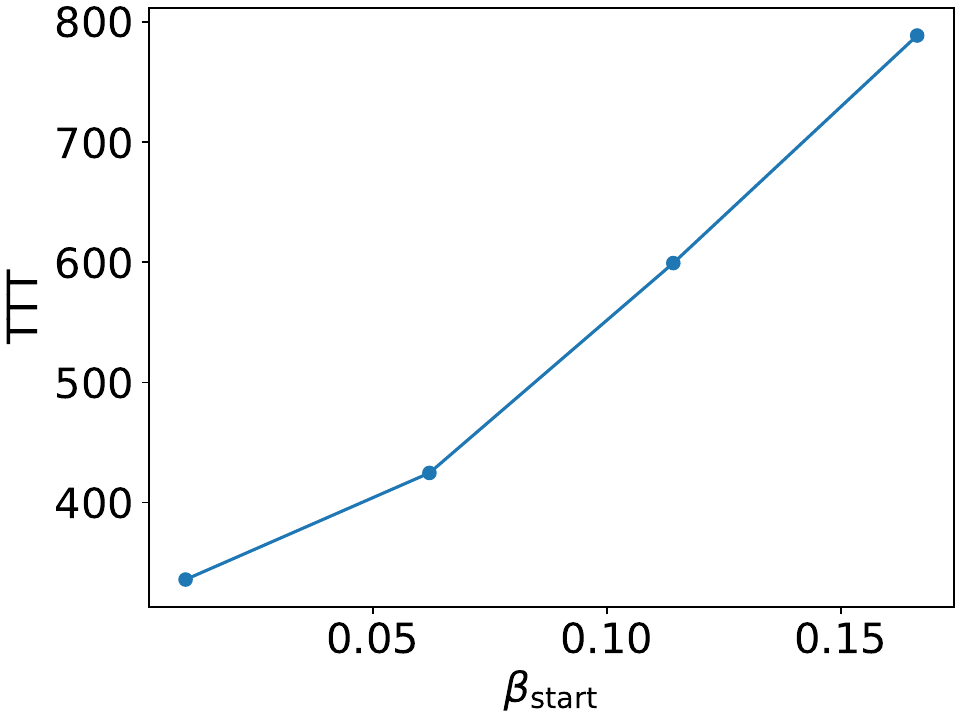}
    \caption{{\bf Performance dependence on the initial coupling strength.} The time-to-target (TTT) of the linear hybrid CAA method using the spin sign method on Beasley instance bqp50-1 as a function of initial coupling strength $\beta_{\text{start}}$. The values of the TTT are obtained using the following fixed parameters: $\alpha=-2.1$, $N_t=1000$, $v_\beta=0.001$ and $\gamma= 0.1$}
    \label{fig:Beta_start_dependence_Beasley}
\end{figure}

To verify if lower values of the initial coupling strength $\beta_\text{start}$ also work better for the Beasley instances, we obtained the TTT of the hybrid CAA method using the spin sign method for several values of $\beta_\text{start}$. In Fig.~\ref{fig:Beta_start_dependence_Beasley}, this relationship is shown for instance bqp50-2, revealing a clear trend. Similar to the MaxCut instances, lower values of $\beta_\text{start}$ result in lower TTT and therefore better performance, illustrating that $\beta_\text{start} = 0.01$ is a good choice. For higher values of $\beta_\text{start}$, the ground state was no longer found during any of the runs, yielding an infinite TTT. Therefore, these data points are not shown in Fig.~\ref{fig:Beta_start_dependence_Beasley}. For all other instances, a similar trend is observed. If the spin sign method is not used, this trend is less universal, but this choice of $\beta_\text{start}$ still results in the best performance for most problem instances.

\end{document}